\documentclass[11pt]{article}
\usepackage{amsfonts, amsmath, amssymb,epsfig}
\usepackage[latin1]{inputenc}
\usepackage[english]{babel}
\usepackage{color}
\usepackage{hyperref}
\newtheorem{thm}{Theorem}[section]
\newtheorem{cor}[thm]{Corollaire}
\newtheorem{lem}[thm]{Lemme}
\newtheorem{pro}[thm]{Proposition}
\newtheorem{dfn}[thm]{Definition}
\newtheorem{rmq}[thm]{Remark}
\newtheorem{expl}[thm]{Exemple}

\def\dessous#1\sous#2{\mathrel{\mathop{\kern0pt#2}\limits_{#1}}}

\newcommand{\0}{/ \! \! \! 0}

\newcommand{\beq}{\begin{eqnarray}}
\newcommand{\eeq}{\end{eqnarray}}
\newcommand{\bpro}{\begin{pro}}
\newcommand{\epro}{\end{pro}}
\newcommand{\blem}{\begin{lem}}
\newcommand{\elem}{\end{lem}}
\newcommand{\bdfn}{\begin{dfn}}
\newcommand{\edfn}{\end{dfn}}
\newcommand{\bcor}{\begin{cor}}
\newcommand{\ecor}{\end{cor}}
\newcommand{\bthm}{\begin{thm}}
\newcommand{\ethm}{\end{thm}}
\newcommand{\bex}{\begin{expl}}
\newcommand{\eex}{\end{expl}}
\newcommand{\brmq}{\begin{rmq}}
\newcommand{\ermq}{\end{rmq}}
\newcommand{\benum}{\begin{enumerate}}
\newcommand{\eenum}{\end{enumerate}}
\newcommand{\bitem}{\begin{itemize}}
\newcommand{\eitem}{\end{itemize}}

\begin{document}

\begin{titlepage}
\begin{flushright}
\end{flushright}
   \begin{center}
     {\ }\vspace{0.5cm}
      {\bf {Density operator approach for Landau problem 
quantum Hamiltonians}}

 \vspace{0.5cm}

 Isiaka Aremua$^{a,b,\dag}$, Mahouton Norbert Hounkonnou$^{b,\ddag}$ and Ezinvi Balo\"{\i}tcha$^{b,\star}$

 \vspace{0.5cm}
 
 $^{a}${\em
Universit\'{e} de Lom\'{e} (UL)}\\
{\em Facult\'{e} Des Sciences  (FDS), D\'{e}partement de Physique} \\
{\em B.P. 1515 Lom\'{e}
TOGO }\\  

       $^{b}${\em  International Chair of Mathematical Physics
 and Applications} \\
 {\em ICMPA-UNESCO Chair}\\
 {\em University of Abomey-Calavi}\\
 {\em 072 B.P. 50 Cotonou, Republic of Benin}\\

 {\em E-mail: {\tt  $^{\dag}$claudisak@gmail.com,  
  $^\ddag$norbert.hounkonnou@cipma.uac.bj,$^\star$ezinvi.baloitcha@cipma.uac.bj.
 \footnote{Corresponding author
  (with copy to hounkonnou@yahoo.fr)}

 }}

    \vspace{2.0cm}
           \today

\abstract{In this work, the definition of the density operator on quantum states in
Hilbert spaces and some of its aspects relevant in thermodynamics and information-theoretical entropy calculations are  given. 
In this framework,  a physical 
model describing an electron in a magnetic
field is investigated. The so-called exotic Landau problem in noncommutative plane is also considered. 
Then,  
 a model related to the  fractional quantum Hall effect is revisited.  
Thanks to the completeness relations verified by the coherent states (CS) in these models, 
the  thermodynamics is discussed by using the diagonal $P$-representation of the
density operator. 
Specifically, the $Q$-Husimi
distribution and the Wehrl entropy
are determined.}

 \end{center}
 
\end{titlepage}

\section{Introduction}
\label{sec:1}
In quantum mechanics, the {  {\it thermal} }
density operator is  {  powerfully} used in order to represent ensembles of {  pure or mixed} quantum states.
See for e.g., \cite{popov, gazbook09}, and references therein.  
For the usual treatment of equilibrium in statistical mechanics using the Gibbs's canonical distribution 
(\cite{popov, pennini}), the    normalized  density operator is  given by 
$ \rho = \frac{1}{Z}e^{-\beta H},$
 where $Z = \mbox{Tr}(e^{-\beta H})$ is the partition function, $\beta = 1/k_{B}T;$  $T $ is the  temperature, and $k_{B}$ 
 the Boltzmann constant which, in SI units, has the value $1.3806503 \times 10^{-23} J/K.$
 The  diagonal expansion of the density operator,  known as the Glauber-Sudarshan 
(GS)-$P$-representation, was introduced independently by Glauber \cite{glauber} and Sudarshan \cite{sudarshan} for the harmonic 
oscillator {coherent states
(CS)}. It is given in GS CS
$|\alpha\rangle$ framework by \cite{glauber, cahill-glauber, brif-ben-aryeh}
{\small{
\begin{equation}
\rho  = \int d^2 \alpha \, \, P(\alpha)|\alpha\rangle \langle \alpha|
\end{equation}
}}
with $P(\alpha)$ a quasiprobability distribution function.

In quantum information, for an ensemble, e.g., of qubits, the density operator was used to describe the 
informational content of the ensemble \cite{popov}. The    quantum-mechanical phase-space distributions of the harmonic 
oscillator CS  were shown to be useful in different situations. See for e.g., \cite{glauber, klauder-skagerstam, pennini}. 
Particularly, in \cite{anderson-halliwell}, the concepts of Husimi distribution \cite{husimi} and Wehrl \cite{wehrl}  entropy, needed in 
the generalized, Fisher, and Shannon informations measures \cite{pennini}, were discussed.   

 In \cite{popov}, the density operator was built for photon-added Barut-Girardello CS 
    in the cases of  pseudoharmonic oscillator and
 generalized hypergeometric thermal CS, with the relevant statistical properties.  
 In \cite{parthasarathy-sridhar}, a $q$-analogue of the diagonal representation of the density matrix, using $q$-boson
 CS, was derived,  and  the $q$-generalization of the density matrix self-reproducing property was discussed. Besides, 
a  Glauber-Sudarshan $P$-representation of the density matrix and relevant issues
related to the properties of the  reproducing kernel  were investigated for a construction of a dual pair of nonlinear CS 
  for a model  obeying a $f$-deformed Heisenberg algebra \cite{aremua-hounk-balo}.  More recently \cite{sodogaetal},  the density matrix of a quantum canonical ideal  gas of a 
 system in thermodynamic equilibrium  was  given in the generalized photon-added associated hypergeometric CS
(GPAH-CS), with a discussion of statistical analysis   in the context of photon-added CS for shape invariant potentials \cite{sodogaetal}.
 
The behavior of an electron in an external magnetic field was extensively studied \cite{landau}-\cite{aremuaetal-HSchm}. This  implied 
a great interest  to other similar physical systems describing, for instance, the quantum Hall effect \cite{girvin}.
In some previous works, this physical model  wasalso  proved to show an interesting application \cite{alietal} 
of the Tomita-Takesaki modular
theory \cite{bratelli}-\cite{tomita}.  
In \cite{ahb}, the density operator  was achieved in the Barut-Girardello CS representation for Landau levels 
of a gas of spinless charged particles, subject to a perpendicular magnetic field confined in a harmonic potential;  the
Husimi distribution  and Wehrl  entropy were investigated.  Recently
 \cite{aremuaetal-HSchm}, 
the Hilbert-Schmidt operators and the Tomita-Takesaki modular theories were recast for noncommutative quantum mechanics formulation, and
the  density matrix formalism related to von Neumann algebras,  displayed by 
the Landau problem \cite{alietal},  was revisited.

 Our present contribution paper  is organized as follows:
\begin{itemize}
\item
First,  we recast the density operator theory  in the
framework of CS by giving  some basic  preliminaries and usual definitions.

\item    
Then,   we apply the density operator approach to a physical model 
describing the motion of a charged particle on the flat plane xy in
the presence of a constant magnetic field along the $z$-axis. First, the study is performed in  
the noncommutative quantum mechanics formalism \cite{scholtzetal}. Next, it is achieved in the context
 of modular theory based on Hilbert-Schmidt operators for which thermal CS were constructed \cite{alietal}.

 \end{itemize}

\section{Preliminaries-Density operator and coherent states}
\label{sec:2}
This section is devoted to some basic facts about the density operator in the CS setting, 
the Wehrl entropy and the Husimi distribution. More details can be found in  
\cite{pennini,wehrl, anderson-halliwell, brif-ben-aryeh, husimi, popov}.  

 \subsection{Coherent states} 
  Coherent states (CS) were introduced for the first time by 
Schr\"{o}dinger \cite{schrodinger} in 1926 for the quantum harmonic oscillator
as the specific quantum state   which has dynamical behavior   similar  to that of the classical harmonic oscillator.
They were rediscovered   by Klauder \cite{klauder}
 in a mathematical physics application,
 and   by  Glauber \cite{glauber} and 
Sudarshan \cite{sudarshan} in the context of quantum optics   at the beginning
of the 1960's. CS are useful in   condensed matter physics, quantum optics, 
quantum field theory,  quantization problems,     quantum information, etc. \cite{klauder-sudarshan}-\cite{gazbook09, agh}. 
Apart from the canonical or harmonic oscillator CS, 
CS  are also generated  
as the lowering operator eigenstate (the so-called CS  of the Barut-Girardello kind) or by applying the displacement operator 
on a ground state 
(Klauder-Perelomov CS) or CS of the Gazeau-Klauder kind, including the nonlinear CS, squeezed states and deformed CS, see  
\cite{klauder-skagerstam,  
ali-antoine-gazeau, dodonov, gazbook09}. In the group theory approach,  they are built as the orbit
under the action of a group representation \cite{perelomov, ali-antoine-gazeau}.

CS  can  be defined over complex domains in the Hilbert space $\mathfrak{H} = span \{\phi_{m}, m \in \mathbb N \}$, 
 which  realizes  at the mathematical side,  the skeleton of quantum theories,  as \cite{ali-antoine-gazeau} 
 {\small{
\begin{equation}{\label{covcs}}
|z\rangle  = (\mathcal{N}(|z|))^{-\frac{1}{2}}\sum_{m=0}^{\infty}\frac{z^{m}}{\sqrt{\rho(m)}}|\phi_{m}\rangle, \qquad z = re^{i\theta}
\end{equation}
}}
where $\{\rho(m)\}_{m=0}^{\infty}$ is a sequence of non-zero positive numbers  chosen so as to ensure the convergence of the sum in a 
non-empty open 
subset $\mathcal D$  of the complex plane, $\mathcal N(|z|)$ is the normalization factor ensuring that $\langle z|z\rangle  = 1$.

The resolution of the identity   is given by 
{\small{
\begin{equation}{\label{rescovcs}}
\int_{\mathcal{D}}|z\rangle \langle z| d\mu = I_{\mathfrak{H}},
\end{equation}
}}
where $d\mu$ is an appropriate chosen measure and $I_{\mathfrak{H}}$ the identity operator on the Hilbert space $\mathfrak H$.

\subsection{The $P$-distribution function}
\label{subsec:2}
The diagonal expansion \cite{brif-ben-aryeh} of the normalized canonical density operator is
{\small{
\begin{equation}\label{thermal03}
\rho  = \int_{\mathcal D} d\mu |z\rangle P(|z|^2) \, \langle z|, \quad \int_{\mathcal D} d\mu \,   P(|z|^2) = 1
\end{equation}
}}
where the $P$-distribution  function  $P(|z|^2)$ satisfying the normalization to unity condition must  be determined.

\subsection{The  $Q$-Husimi function or distribution}
\label{subsec:3}

The Husimi distribution \cite{husimi} as a function in the phase space, in general viewed as Gaussian  smoothing of the Wigner function, 
is derived by use of the expected value of the density operator in a basis of CS \cite{klauder-skagerstam}.

Taking the   normalized  density operator  $\rho$ given in the basis $\{|n\rangle\}_{n=0}^{\infty}$ by  
{\small{
\begin{equation}
\rho = \frac{1}{Z(\beta)}\sum_{n=0}^{\infty}e^{-\beta E_n}|n\rangle \, \langle n|
\end{equation}
}}
with $Z(\beta)$ the partition function,  the $Q$-Husimi function or distribution\cite{husimi}  is provided as 
{{\small
\begin{equation}
Q(|z|^2) =  \langle z| \rho |z\rangle, \quad \mbox{with} \quad \int_{\mathcal D} d\mu \; Q(|z|^2) = 1
\end{equation}
}}

assuring the normalization to unity of the $Q$-Husimi function.

The normalization of the density operator in the CS (\ref{covcs}) $\{|z\rangle\}$ basis  leads to
{\small{
\begin{equation}\label{thermal02}
\mbox{Tr} \rho  =  \int_{\mathcal D}  d\mu  \, \langle z|  \rho |z\rangle =  1.
\end{equation}
}}

\subsection{The Wehrl entropy}
\label{subsec:4}

The Wehrl entropy or the ``classical'' entropy associated with a quantum system is the entropy of the probability distribution
in phase-space, corresponding to the Husimi $Q$-function in terms of CS \cite{wehrl} 
(see also \cite{anderson-halliwell, lieb, pennini,popov} and references therein). 
It is of great importance for the measure of localization in the phase-space  \cite{wehrl} and constitutes a powerful tool 
in statistical physics. 
 It is   given by \cite{wehrl, pennini} 
{\small{
\begin{equation}\label{wehrl00}
W: = -\int \frac{dx dp}{2\pi \hbar} \mu(x,p) \ln \mu(x,p)
\end{equation}
}}

where $\mu(x,p)$ in the CS (\ref{covcs}) basis is such that:  $\mu(x,p) =  \langle z|\rho|z \rangle $ being 
the $Q$-Husimi function or distribution corresponding to ``semi-classical'' phase-space distribution function associated to the 
density matrix $\rho$, with $z=\sqrt{\frac{m\omega}{2\hbar}}x + i\sqrt{\frac{\hbar m \omega}{2}}p.$ 

From (\ref{rescovcs}), taking $\mathcal{D} = \mathbb C, d\mu = \frac{d^2 z}{\pi}, $ we have

{\small{
\begin{equation}
I_{\mathfrak{H}} = \int_{\mathbb C}|z\rangle \langle z| \frac{d^2 z}{\pi} = \int \frac{dx dp}{2\pi \hbar}|x, p\rangle \langle x, p| 
\end{equation}
}}

such that (\ref{wehrl00})  takes the form

{\small{
\begin{equation}
W = \int_{\mathbb C}\frac{d^2 z}{\pi}  \langle z|\rho|z \rangle \ln( \langle z|\rho|z \rangle).
\end{equation}
}}

Both  $Q$-Husimi function and Wehrl entropy are used in  information-theoretical entropy of 
some quantum oscillators, and also in Fisher's and 
Shannon information measures  in statistical mechanics\cite{pennini, popov}.

\section{Density operator approach in CS for the exotic Landau problem}

 The Landau problem  \cite{landau}-\cite{dodonov1}-\cite{aremuaetal-HSchm}
is related to the motion of a charged particle on the flat plane xy in
the presence of a constant magnetic field along the $z$-axis. 
In metals,  the electrons occupy many Landau levels   
 $E_{n} = \hbar \omega_{c} (n+\frac{1}{2})$, each level being infinitely 
degenerate, with $\omega_{c} = eB/Mc$, the cyclotron frequency, which correspond to the kinetic
energy levels of electrons, and are those of the one-dimensional harmonic oscillator. Here, we deal 
with the Landau exotic problem \cite{horvathy}.  We use   the formalism developed in
\cite{a-hk, scholtzetal} to construct CS for this model. From this setup, we
derive the related density operator.

\subsection{The model}
Let us first make a brief review  of the main features of \textquotedblleft exotic \textquotedblright particles. 
An exotic particle is a particle
moving in a planar electromagnetic field  ${\bf E }$ and ${\bf B }$ [assumed static for simplicity]  described
by the equations \cite{horvathy}
{\small{
\begin{equation}\label{equa00}
M^{*}\dot{x}^{i} = p^{i} - M e \theta\varepsilon^{ij}E^{j}, \qquad \dot{p}^{i} = e B \varepsilon^{ij}\dot{x}^{j} + e E^{i}
\end{equation}
}}
with $\varepsilon^{ij}$  the components of the
antisymmetric tensor normalized by $\varepsilon^{12} =1$, 
where $M, e$ and $\theta$ are the mass, charge and noncommutative parameter, respectively, and $M^{*} = (1-eB\theta)M$
 is the effective mass. The equations (\ref{equa00}) derive from the symplectic form and Hamiltonian \cite{horvathy},
 {\small{
\begin{equation}{\label{sol011}}
\Omega = dp^{i}\wedge dx^{i} + \frac{\theta}{2}\varepsilon^{ij} dp^{i}\wedge dp^{j} + \frac{eB}{2}\varepsilon^{ij} dx^{i}\wedge dx^{j}, 
\quad  \mathcal H = \frac{{\bf p}^{2}}{2M} + V(x)
\end{equation}
}}

respectively, through the \textquotedblleft exotic\textquotedblright Poisson brackets

{\small{
\begin{equation}\label{pbrack}
\{x^{i}, x^{j}\}  = \frac{\theta}{1-eB\theta}\varepsilon^{ij}, 
\qquad \{x^{i}, p^{j}\}  = \frac{\delta^{ij}}{1-eB\theta}, \qquad \{p^{i}, p^{j}\} = \frac{eB}{1-eB\theta}\varepsilon^{ij}
\end{equation}
}}

where the non-critical regime, $eB\theta \neq 1$, is assumed.  When the magnetic field takes the critical value
{\small{
\begin{equation}
B = B_c = \frac{1}{e\theta},
\end{equation}
}}
the system becomes singular: the determinant of the symplectic matrix is given by $\mbox{det}(\Omega_{\alpha\beta}) = (M^{*}/M)^{2} =0$, 
and consistency requires the Hall law,

{\small{
\begin{equation}
p^{i} = M e \theta\varepsilon^{ij}E^{j}, \qquad \dot{x}^{i} = \varepsilon^{ij}\frac{E^{j}}{B},
\end{equation}
}}
to be satisfied \cite{horvathy}.

Consider the physical Hamiltonian, given in (\ref{sol011}),  related to the exotic Landau problem  \cite{horvathy}, 
where the \textquotedblleft exotic \textquotedblright Poisson brackets (\ref{pbrack}) with $i, j = 1, 2$ hold. 
{The Hamiltonian (\ref{sol011}),  writes  with chiral coordinates  ${\bf X_{\pm}}$  which describe the  
system and  given by \cite{alvarez, horvathy}:
{\small{
\begin{equation}\label{pbrack00}
\{X^{i}_{+}, X^{j}_{+} \} = -\frac{1}{eB}\varepsilon^{ij}, \quad  \{X^{i}_{+}, X^{j}_{-} \} = 0, \quad 
\{X^{i}_{-}, X^{j}_{-} \} = \frac{1}{eB(1-eB\theta)}\varepsilon^{ij}
\end{equation}
}}
where the  extension of the original coordinates can be  performed as  
{\small{
\begin{equation}
X^{i}_{+} = x^i  + \frac{1}{eB}\varepsilon^{ij}\lambda^{+}_{j},  \qquad 
X^{i}_{-} = x^i  - \frac{1}{eB}\varepsilon^{ij}\lambda^{-}_{j}
\end{equation}
}}
with $ \lambda^{\pm}_{i}= p_i - m_iv_i $ measuring the difference between the canonical $(p_i )$ and the mechanical $(mv_i )$ momenta  
\cite{horvathy} and 
satisfying 
{\small{
\begin{equation}
\{\lambda^{+}_{i}, \lambda^{+}_{j} \} = \frac{eB}{1-eB\theta} \varepsilon^{ij}, 
\qquad \{\lambda^{-}_{i}, \lambda^{-}_{j} \} = -eB  \varepsilon^{ij}.
\end{equation} 
}}

Then, the Hamiltonian (\ref{sol011})   splits  into 
two uncoupled systems  both with 2d phase spaces as follows  \cite{alvarez, horvathy}:
{\small{
\begin{equation}{\label{sol012}}
\mathcal H = \mathcal H_{+} + \mathcal H_{-} = \left\{-e {\bf E } . {\bf X_{+}} \right\}  + \left\{\frac{(eB)^{2}}{2M} {\bf X^{2}_{-}} 
- e {\bf E } . {\bf X_{-}} \right \}
\end{equation}
}}
where the symplectic form writes as
{\small{
\begin{equation}\label{form}
\Omega = \Omega_{+} + \Omega_{-} =  \left\{\frac{eB}{2}(\varepsilon^{ij} dX^{i}_{+}\wedge dX^{j}_{+})\right\}  - 
\left\{(1-eB\theta)\frac{eB}{2}(\varepsilon^{ij} dX^{i}_{-} \wedge dX^{j}_{-}) \right\}. 
\end{equation}
}}

}


\subsection{The quantum Hamiltonian}\label{quanthamilt}

We investigate the quantum Hamiltonian in  the purely magnetic case
$E = 0$ \cite{a-hk-landauexotic}.  Let us assume that $V$ (but not $B$)
vanishes, so that  the Hamiltonian (\ref{sol011}) reduces to

{\small{
\begin{equation}
\mathcal H = \frac{{\bf p}^{2}}{2M}
\end{equation}
}}
where we deal with the non-critical regime, $eB\theta \neq 1$.

The commutators associated to the relations (\ref{pbrack}) are given by

{\small{
\begin{equation}
[ x^{i},  x^{j}]  = \frac{i\theta}{1-eB\theta}\varepsilon^{ij}, 
\qquad [ x^{i},  p^{j}]  = \frac{i\delta^{ij}}{1-eB\theta}, \qquad [ p^{i},  p^{j}] = i\frac{eB}{1-eB\theta}\varepsilon^{ij}.
\end{equation}
}}

In order to quantize  the Hamiltonian  physical model, we consider, by assuming the relation $1-eB\theta > 0$, 
 a more convenient system of coordinates by   
introducing  a set of chiral 
complex coordinates  
${\mathcal Z_{\pm}}$ defined, by setting
{\small{
\begin{equation}
\mathcal X^{i}_{+} = \sqrt{\frac{eB\theta}{1-eB\theta}}X^{i}_{+}, \qquad  
\mathcal X^{i}_{-} = \sqrt{eB\theta}X^{i}_{-},\qquad
i=1,2,
\end{equation}
}}
 as

 {\small{
\begin{equation}
\mathcal Z_{+} = \sqrt{\frac{1-eB\theta}{2\theta}}[\mathcal X^{1}_{+} - i \mathcal X^{2}_{+}], \qquad \mathcal Z_{-} = \sqrt{\frac{1-eB\theta}{2\theta}}
[\mathcal X^{1}_{-} + i \mathcal X^{2}_{-}].
\end{equation}
}}

Then, from the relations (\ref{pbrack00}) satisfied by the set $\{X^{i}_{\pm}, X^{j}_{\pm}, i,j=1,2 \}$,  it follows

{\small{
\begin{equation}\label{pbrack03}
\{\mathcal Z_{+}, \bar{\mathcal Z}_{+}\} = -i , \quad \{\mathcal Z_{+}, \bar{\mathcal Z}_{-}\} = 0 = \{\bar{\mathcal Z}_{+},  \mathcal Z_{-}\}, \quad  
\{\mathcal Z_{-}, \bar{\mathcal Z}_{-}\} = -i.
\end{equation}
}}

Next, denote by $\{\hat{\mathcal Z}_{\pm}, \hat{\mathcal Z}^{\dag}_{\pm}\}$, where $ \mathcal Z^{\dag}_+ \equiv \bar{\mathcal Z}_+$, 
 the corresponding operators of the chiral coordinates $\{\mathcal Z_{\pm}, \bar{\mathcal Z}_{\pm}\}$. Then, the classical structure (\ref{pbrack03}) is 
replaced by commutators  as follows:

{\small{
\begin{equation}
\{\mathcal A, \mathcal B\}  \quad \rightarrow \quad   \frac{1}{i} [\hat{\mathcal A}, \hat{\mathcal B}], \qquad \hbar =1
\end{equation}
}}

where $\hat{\mathcal A}, \hat{\mathcal B}$ denote the quantized variables such that

{\small{
\begin{equation}\label{opchi01}
[\hat{\mathcal Z}_{+}, \hat{\mathcal Z}^{\dag}_{+}] = 1 = [\hat{\mathcal Z}_{-}, \hat{\mathcal Z}^{\dag}_{-}], \quad 
[\hat{\mathcal Z}_{+}, \hat{\mathcal Z}^{\dag}_{-}] = 0 = [\hat{\mathcal Z}_{-}, \hat{\mathcal Z}^{\dag}_{+}], \quad [\hat{\mathcal Z}_{+}, \hat{\mathcal Z}_{-}] = 0.
\end{equation}  
}}

The relations (\ref{opchi01}) displaying that $\{\hat{\mathcal Z}_{\pm}, \hat{\mathcal Z}^{\dag}_{\pm}\}$ form an irreducible set of operators on the 
chiral boson Fock spaces $\mathcal F_{\pm} = \{|n_{\pm}\rangle \}^{\infty}_{{n_\pm}=0}$,  
 suggest to identify 
the system $\{\hat{\mathcal Z}_{\pm}, \hat{\mathcal Z}^{\dag}_{\pm}\}$   with the annihilation and creation operators   relevant in 
the formalism of noncommutative quantum mechanics developed in \cite{scholtzetal} and acting  on the states 
 $|n_{+},  n_{-}) =
|n_{+} \rangle \langle n_{-}|,  \;  n_{\pm} = 0, 1, 2, \dots,$  as

{\small{
\begin{equation}\label{opchi02}
\hat{\mathcal Z}_{+}| n_{+},  n_{-})   =   \sqrt{n_{+}}| n_{+}-1, n_{-})
\quad \quad  \hat{\mathcal Z}^{\dag}_{+}|n_{+}, n_{-}) = \sqrt{ n_{+} + 1 }| n_{+}+1,  n_{-}),
\end{equation}
\begin{equation}\label{opchi03}
\hat{\mathcal Z}_{-}| n_{+},  n_{-})  =   
\sqrt{ n_{-}}|n_{+}, n_{-}-1)
\quad \quad  \hat{\mathcal Z}^{\dag}_{-}|n_{+}, n_{-}) =  \sqrt{n_{-} + 1 } | n_{+},  n_{-}+1).
\end{equation}
}}

We have

{\small{
\begin{equation}\label{eig00}
|n_{+}, n_{-})
= \frac{1}{\sqrt{n_{+} ! n_{-} !}}\left(\hat{\mathcal Z}^{\dag}_{+}\right)^{n_{+}}
\left(\hat{\mathcal Z}^{\dag}_{-}\right)^{n_{-}}|0\rangle \langle 0|
\end{equation}
}}
where $\hat{\mathcal Z}^{\dag}_{-}$ may have an action on the right by $\hat{\mathcal Z}_{-}$ on $|0\rangle \langle 0|$. 
  $||| n_{+}, n_{-})|| = 1$ and  $|0\rangle \langle 0|$  
stands for the vacuum state on $\mathcal H_{q}$ i.e.  the space of Hilbert-Schmidt operators
acting on the noncommutative configuration (Hilbert) space $\mathcal H_{c}$ 
(isomorphic to the boson Fock space).
$\mathcal H_{q}$ is defined as:
{\small{
\begin{equation}
\mathcal H_{q} = \left\{\psi(\hat x_{1}, \hat x_{2}): \psi(\hat x_{1}, \hat x_{2}) \in \mathcal B(\mathcal H_{c}),\,
tr_{c}(\psi(\hat x_{1}, \hat x_{2})^{\dag}, \psi(\hat x_{1}, \hat x_{2}))
 < \infty \right \}
\end{equation}
}}
$\hat x_{1}, \hat x_{2}$ being the coordinates of noncommutative configuration space. $\mathcal H_{q}$ is  
endowed with the following inner product
{\small{
\begin{equation}
(\psi(\hat x_{1}, \hat x_{2}), \phi(\hat x_{1}, \hat x_{2})) = tr_{c}(\psi(\hat x_{1}, \hat x_{2})^{\dag}, \phi(\hat x_{1}, \hat x_{2}))
\end{equation}
}}
where $tr_{c}$ stands for the trace over $\mathcal H_{c}$, 
$\mathcal B(\mathcal H_{c})$  being the set of bounded operators on $\mathcal H_{c}$.

Then, the eigenvalues of the Hamiltonian $\hat H $ with the following chiral decomposition
{\small{
\begin{equation}\label{quantumhamilt}
\hat H = \hat H_{+}  \otimes I_{\mathcal F_{-}} + I_{\mathcal F_{+}}\otimes \hat H_{-} 
\end{equation}
}}
where $I_{\mathcal F_{\pm}}$ are the identity operators on the chiral boson Fock spaces 
$\mathcal F_{\pm} = \{|n_{\pm}\rangle \}^{\infty}_{{n_\pm}=0}$, respectively, are given by

{\small{
\begin{equation}\label{eig01}
E_{n_{+},  n_{-}} = \hbar \omega_{c}\left(n_{+} + \frac{1}{2}\right) + \hbar \omega^{*}_{c}\left(n_{-} + \frac{1}{2}\right) 
\end{equation}
}}
where $\omega_{c} = \frac{eB}{M}, $ $\omega^{*}_{c} = \frac{eB}{M^{*}}, M^{*} = (1-eB\theta)M$, with associated eigenstates given
in (\ref{eig00}).

With the help of the operators $\{\hat{\mathcal Z}_{\pm}, \hat{\mathcal Z}^{\dag}_{\pm}\}$ provided in  (\ref{opchi01})-(\ref{eig00}),    
since there exists a set of eigenstates $|z_{\pm}\rangle $ satisfying 
{\small{
\begin{equation}
\hat{\mathcal Z}_{\pm}|z_{\pm}\rangle  = z_{\pm}|z_{\pm}\rangle, \qquad \langle z_{\pm}|\hat{\mathcal Z}^{\dag}_{\pm} = \langle z_{\pm}|\bar z_{\pm}
\end{equation}
}}
having complex eigenvalues $z_{\pm}$ with   
{\small{
\begin{equation}\label{cohst00}
|z_{\pm}\rangle  = e^{-\frac{|z_{\pm}|^2}{2}}e^{\{z_{\pm} \hat{\mathcal Z}^{\dag}_{\pm}\}} |0\rangle 
\end{equation}
}}

given in terms of the chiral Fock basis and provided by  the  Baker-Campbell-Hausdorff identity 
\begin{equation} 
e^{\{z_{\pm} \hat{\mathcal Z}^{\dag}_{\pm} - \bar z_{\pm} \hat{\mathcal Z}_{\pm}\}} = e^{-\frac{|z_{\pm}|^2}{2}}e^{\{z_{\pm} \hat{\mathcal Z}^{\dag}_{\pm}\}}
e^{\{-\bar z_{\pm} \hat{\mathcal Z}_{\pm}\}},
\end{equation}
the  CS of the noncommutative plane related to the quantum Hamiltonian (\ref{quantumhamilt})  are given by
{\small{
\begin{equation}\label{vect00}
|z_{\pm} ) = |z_{+}\rangle  \langle z_{-}|= 
e^{-\frac{1}{2}(|z_{+}|^{2} + |z_{-}|^{2})}
\sum_{{n}_{+}, {n}_{-} = 0}^{\infty}
\frac{z_{+}^{{n}_{+}}\bar z_{-}^{{n}_{-}}}{\sqrt{{n}_{+} !{n}_{-} !}}
|{n}_{+}\rangle \langle {n}_{-}|. 
\end{equation}
}}

The identity operator $\mathbb I_{q}$  on $ \mathcal H_{q}$ writes in terms of the states $|n_+, n_-)$   as follows:
{\small{
\begin{equation}
\mathbb I_{q}
 =  \sum_{n_+, n_- = 0}^{\infty}|n_{+} \rangle \langle n_{-}|
|n_{-}\rangle\langle n_{+}|. 
\end{equation}
}}

\bpro

{\small{
The CS $|z_{\pm})$ satisfy the resolution of the identity \cite{a-hk}
\begin{equation}{\label{resolv}}
\frac{1}{\pi^{2}}\int_{\mathbb C^{2}}|z_{\pm})
(z_{\pm}|d^{2}z_{+}d^{2}z_{-} \equiv \mathbb I_{q}
\end{equation}
}}

where the identity operator on $ \mathcal H_{q}$ is  given by \cite{scholtzetal}

{\small{
\begin{equation}{\label{resolv01}}
\mathbb I_{q} = \frac{1}{\pi}\int_{\mathbb C}dzd\bar{z}|z)e^{\overleftarrow{\partial_{\bar z}}\overrightarrow{\partial_{z}}} (z|.
\end{equation}
}}
\epro
 
{

{\bf Proof.}   See \cite{a-hk}.

$\hfill{\square}$

\subsection{Statistical properties}

In this section, we will carry out a  discussion on the statistical properties of the CS $|z_{+}, z_{-})$
for the Landau exotic problem. 
We 
use the results issued from the paragraph \ref{quanthamilt} as   key ingredients  to construct the diagonal 
$P$-representation, derive diagonal elements of the density 
operator $\rho$    characterizing the probability distribution on the states of a physical system,  and 
examine its physical   properties (see
for e.g. \cite{ahb} and references listed therein).  

Considering that the quantum system obeys  the  canonical distribution, let us take the partition function $Z$ as that of 
a composite system made of two independent systems such that is the product of the partition functions  of the components, i.e.    
$Z = Z_+  Z_-$. 
\bpro

The  diagonal elements   of the normalized density operator 
$\rho = \frac{1}{Z} e^{-\beta \hat H}$ in the CS 
$|z_{+}, z_{-}) $ representation, also known as the $Q$-distribution function or the $Q$-Husimi distribution
\cite{mandel-wolf}, are  derived as
{\small{
\begin{eqnarray}\label{density00}
(z_{+}, z_{-}|\rho|z_{+}, z_{-}) 
&=& \frac{1}{\bar{n}+1} e^{-\frac{1}{\bar{n}+1}|z_+|^2} \times \frac{1}{{\bar{n}}^{*}+1} e^{-\frac{1}{{\bar{n}}^{*}+1}|z_+|^2}\cr
&=& Q(|z_+|^2)Q(|z_-|^2)
\end{eqnarray}
}} 
with $\bar{n} = \left[e^{\beta \hbar \omega_c }-1\right]^{-1}$ and ${\bar{n}}^{*} =\left[e^{\beta \hbar \omega^{*}_c }-1\right]^{-1}$  
being the corresponding thermal expectation values of the number operator (the Bose-Einstein distribution functions for oscillators with 
angular frequencies  $\omega_c$ and $\omega^{*}_c, $ respectively) or the thermal mean occupancy for harmonic oscillators with the angular
frequencies  $\omega_c$ and $\omega^{*}_c, $ respectively. 
\epro

The quantity $|(n_+, n_-|z_{+}, z_{-})|^2$ is such that
{\small{
\begin{equation}\label{distrib}
|(n_+, n_-|z_{+}, z_{-})|^2 =  e^{- |z_+|^{2} }\frac{|z_+|^{2n_+}}{n_+ !} e^{- |z_-|^{2} }\frac{|z_-|^{2n_-}}{n_- !}
\end{equation}
}}
displaying that the CS $|z_{+}, z_{-})$ obey the photon-number Poisson distribution corresponding to a Mandel parameter 
$\mathcal Q = 0$ \cite{mandel-wolf}.  The
 right-hand side of (\ref{density00}) corresponds to the product of two harmonic oscillators
 Husimi distributions.


{\bf Proof.}  We have
{\small
\begin{eqnarray}
(z_{+}, z_{-}|\rho|z_{+}, z_{-}) &=& \frac{1}{Z}\sum_{n_+,n_-=0}^{\infty}e^{-\beta H}|(n_+, n_-|z_{+}, z_{-})|^2 \cr
&=& \frac{1}{Z}\sum_{n_+,n_-=0}^{\infty}e^{-\beta H_+}e^{-\beta H_-}
e^{- |z_+|^{2} }\frac{|z_+|^{2n_+}}{n_+ !} e^{- |z_-|^{2} }\frac{|z_-|^{2n_-}}{n_- !}, \,\,
\mbox{with} \quad  Z =  Z_+  Z_-\cr
 &=& \left\{\frac{1}{Z_+}e^{-\frac{\beta\hbar  \omega_c}{2}}e^{- |z_+|^{2}}\left[e^{e^{-\beta\hbar \omega_c }|z_+|^{2}}\right]\right\}\left\{\frac{1}{Z_-}e^{-\frac{\beta\hbar  \omega^{*}_c}{2}}e^{- |z_-|^{2}}
 \left[e^{e^{-\beta\hbar \omega^{*}_c}|z_-|^{2}}\right]\right\}\nonumber \\
\end{eqnarray}
}
where
{\small{
\begin{equation}
\frac{1}{Z_+} = \left[\frac{e^{-\frac{\beta\hbar  \omega_c}{2}}}{1-e^{-\beta\hbar  \omega_c}}\right]^{-1},
\qquad \mbox{and} \qquad \frac{1}{Z_-} = 
\left[\frac{e^{-\frac{\beta\hbar  \omega^{*}_c}{2}}}{1-e^{-\beta\hbar \omega^{*}_c}}\right]^{-1}.
\end{equation}
}}

Thereby, 
{\small{
\begin{eqnarray}\label{husimidistr00}
(z_{+}, z_{-}|\rho|z_{+}, z_{-})
&=&\left[1-e^{- \beta \hbar \omega_c }\right]e^{- (1-e^{- \beta \hbar \omega_c })|z_+|^2} \times 
\left[1-e^{- \beta \hbar \omega^{*}_c }\right]e^{- (1-e^{- \beta \hbar \omega^{*}_c })|z_-|^2}\cr
&=& Q(|z_+|^2)Q(|z_-|^2).
\end{eqnarray}
}}
$\hfill{\square}$

The variables changes $r_+ = \left[1-e^{- \beta \hbar \omega_c }\right]^{1/2}|z_+|$ and $r_- =
 \left[1-e^{- \beta \hbar \omega^{*}_c }\right]^{1/2}|z_-|$ 
 with $\frac{d^2 z}{\pi}  =  rdr \frac{d\varphi}{\pi}, \, r \in [0, \infty), \varphi 
\in (0, 2\pi]
$ and  the help of the resolution of the identity (\ref{resolv}), lead to
{\small{
\begin{equation}
 Tr \rho  = \frac{1}{\pi^2}\int_{\mathbb C^{2}}   
d^{2}z_{+}d^{2}z_{-}(z_{+}, z_{-}|\rho|z_{+}, z_{-})  =  1,  
\end{equation}
}}
where we have used the following integral 
{\small{
\begin{equation}
\int_{0}^{\infty} \frac{1}{n_{\pm} !}  2r^{2n_{\pm} +1}_{\pm}dr_{\pm} e^{-r^2_{\pm}} = 1, 
\end{equation}
}}

ensuring that the normalization condition of the density matrix is accomplished.

According to the normalized density operator expression 
{\small{
\begin{eqnarray}
 \rho 
&=&\left\{\frac{e^{-\frac{\beta\hbar  \omega_c}{2}}e^{-\frac{\beta\hbar  \omega^{*}_c}{2}}}{Z_+ Z_-}
\right\}\left\{\sum_{n_+=0}^{\infty}e^{-\beta\hbar \omega_c n_+ }\sum_{n_-=0}^{\infty}
e^{-\beta\hbar \omega^{*}_c n_-}\right\}|n_+, n_-)(n_+, n_-|
\end{eqnarray}
}}
we have
{\small{
\begin{eqnarray}\label{csdensmatrixel00}
(n_+, n_-|{\rho}|n_+, n_-)
&=& \frac{1}{\bar{n}+1}\left(\frac{\bar{n}}{\bar{n}+1}\right)^{n_+} \times 
 \frac{1}{{\bar{n}}^{*}+1}\left(\frac{{\bar{n}}^{*}}{{\bar{n}}^{*}+1}\right)^{n_-}.
\end{eqnarray}
}}

The diagonal expansion of the density operator $\rho$ in the CS $|z_{+}, z_{-})$  is given   with the help of the probability 
$P(|z_+|^2, |z_-|^2)$ as follows:
{\small{
\begin{equation}\label{density}
\rho = \frac{1}{\pi^2}\int_{\mathbb C^{2}}   
d^{2}z_{+}d^{2}z_{-} P(|z_+|^2, |z_-|^2)|z_{+},  z_{-})  (z_{+},   z_{-}|.
\end{equation}
}}

For the Glauber CS $|\alpha \rangle $ of the harmonic oscillator the  expansion (\ref{density}) is called the Glauber-Sudarshan 
$P$-representation 
 of the density operator \cite{cahill-glauber, gazbook09}.

\bpro

The $P$-distribution function   $P(|z_+|^2, |z_-|^2) := P(|z_+|^2)P(|z_-|^2), $ is given,  
by taking $n_{+} = s_{+}-1, n_{-} = s_{-}-1$ and using the resolution of the identity (\ref{resolv}),  as
{\small
\begin{eqnarray}
P(|z_+|^2, |z_-|^2) &=&  \frac{1}{\bar{n}}
\frac{G_{0,1}^{1,0}\left(\left.\frac{{\bar{n}} +1}{\bar{n}} |z_+|^2\right|  \begin{array}{rcl} /;& & / \\0, / & ;   /& \end{array}\right)}
{G_{0,1}^{1,0}\left(\left. |z_+|^2\right|  \begin{array}{rcl} /;& & / \\0, / & ;  /& \end{array}\right)}\frac{1}{{\bar{n}}^{*}}
\frac{G_{0,1}^{1,0}\left(\left.\frac{{{\bar{n}}^{*}} +1}{{\bar{n}}^{*}} |z_-|^2\right|  \begin{array}{rcl} /;& & / \\0, / & ;   /&
\end{array}\right)}
{G_{0,1}^{1,0}\left(\left. |z_-|^2\right|  \begin{array}{rcl} /;& & / \\0, / & ;  /& \end{array}\right)}\cr
&=& \frac{1}{\bar{n}}e^{-\frac{1}{\bar{n}}|z_+|^2}\frac{1}{{\bar{n}}^{*}}e^{-\frac{1}{\bar{n}^{*}}|z_-|^2}
\end{eqnarray}
}
where the  following Meijer's G-function and  Mellin inversion theorem \cite{marichev, prudnikov}
{\small
\begin{eqnarray}
\int_{0}^{\infty}dx x^{s-1}
{G_{p,q}^{m,n}\left(\left. \alpha x\right| 
\begin{array}{rcl} a_{1}, \dots, a_{n};a_{n+1}, \dots, a_{p}\\b_{1},\dots,b_{m};b_{m+1},\dots,b_{q}& \end{array}\right)}
=\frac{1}{\alpha^s}\frac{\prod_{j=1}^{m}\Gamma(b_j+s)}{\prod_{j=m+1}^{q}\Gamma(1-b_j-s)}
\frac{\prod_{j=1}^{n}\Gamma(1-a_j-s)}{\prod_{j=n+1}^{p}\Gamma(a_j+s)} \nonumber
\end{eqnarray}
}

have been used. 

\epro

{\bf Proof.}  Starting from (\ref{density}),  
and using the results (\ref{csdensmatrixel00}) and (\ref{distrib}) together,  we obtain, by setting
{\small
\begin{equation}
\overline{P}(|z_+|^2, |z_-|^2) = P(|z_+|^2, |z_-|^2)G_{0,1}^{1,0}\left(\left. |z_+|^2\right| 
\begin{array}{rcl} /;& & / \\0, / & ;  /& \end{array}\right),
\end{equation}
}

{\small
\begin{eqnarray}
\frac{n_+ !}{\bar{n}+1}\left(\frac{\bar{n}}{\bar{n}+1}\right)^{n_+} \times 
 \frac{n_- ! }{{\bar{n}}^{*}+1}\left(\frac{{\bar{n}}^{*}}{{\bar{n}}^{*}+1}\right)^{n_-} =  
\int_{\mathbb C^{2}}   
\frac{d^{2}z_{+}d^{2}z_{-}}{4\pi^2}\overline{P}(|z_+|^2, |z_-|^2) \left\{|z_+|^{2n_+}|z_-|^{2n_-}\right\}.
\end{eqnarray}
}

Then, taking $n_{+} = s_{+}-1, n_{-} = s_{-}-1, $ we get, after performing the angular integration,  

{\small
\begin{eqnarray}
P(|z_+|^2, |z_-|^2) &=&  \frac{1}{\bar{n}}
\frac{G_{0,1}^{1,0}\left(\left.\frac{{\bar{n}} +1}{\bar{n}} |z_+|^2\right|  \begin{array}{rcl} /;& & / \\0, / & ;   /& \end{array}\right)}
{G_{0,1}^{1,0}\left(\left. |z_+|^2\right|  \begin{array}{rcl} /;& & / \\0, / & ;  /& \end{array}\right)}\frac{1}{{\bar{n}}^{*}}
\frac{G_{0,1}^{1,0}\left(\left.\frac{{{\bar{n}}^{*}} +1}{{\bar{n}}^{*}} |z_-|^2\right|  \begin{array}{rcl} /;& & / \\0, / & ;   /& \end{array}\right)}
{G_{0,1}^{1,0}\left(\left. |z_-|^2\right|  \begin{array}{rcl} /;& & / \\0, / & ;  /& \end{array}\right)}. \nonumber\\
\end{eqnarray}
}

Thus, using the connection between the generalized hypergeometric function and the Meijer's G-function \cite{mathai, popov} leads to
{\small
\begin{eqnarray}
P(|z_+|^2, |z_-|^2)
&=& \frac{1}{\bar{n}} \frac{_{0}F_{0}(; \, ;\, -\frac{{\bar{n}} +1}{\bar{n}}|z_+|^2)}{_{0}F_{0}(; \, ;\, -|z_+|^2)}\times 
\frac{1}{{\bar{n}}^{*}} \frac{_{0}F_{0}(; \, ;\, -\frac{{{\bar{n}}^{*}} +1}{{\bar{n}}^{*}}|z_-|^2)}{_{0}F_{0}(; \, ;\, -|z_-|^2)}\cr
P(|z_+|^2, |z_-|^2) &=& \frac{1}{\bar{n}}e^{-\frac{1}{\bar{n}}|z_+|^2}\frac{1}{{\bar{n}}^{*}}e^{-\frac{1}{\bar{n}^{*}}|z_-|^2}.
\end{eqnarray}
}
$\hfill{\square}$

\bpro

The related Wehrl entropy given by

{\small{
\begin{equation}
W = -\frac{1}{\pi}\left\{-\frac{1}{\pi}\int_{\mathbb C^{2}} (z_{+}, z_{-}|\rho|z_{+}, z_{-}) 
\ln \left\{(z_{+}, z_{-}|\rho|z_{+}, z_{-})\right\}d^{2}z_{+}d^{2}z_{-}\right\} 
\end{equation}
}}
is obtained as 
{\small{ 
\begin{equation}
W = \left[1-\ln \left(1-e^{- \beta \hbar \omega_c }\right)\right]\left[1-\ln \left(1-e^{- \beta \hbar \omega^{*}_c }\right)\right].
\end{equation} 
}}
\epro

{\bf Proof.}  From the Husimi distribution (\ref{husimidistr00}), we get 
{\small
\begin{eqnarray}
W&=& -\frac{1}{\pi}\left\{-\frac{1}{\pi}\int_{\mathbb C^{2}} (z_{+}, z_{-}|\rho|z_{+}, z_{-}) 
\ln \left\{(z_{+}, z_{-}|\rho|z_{+}, z_{-})\right\}d^{2}z_{+}d^{2}z_{-}\right\} \cr
&=& \left\{-\frac{1}{\pi}\int_{\mathbb C}Q(|z_+|^2) \ln\left(Q(|z_+|^2)\right) d^2 z_+ \right\}\left\{-\frac{1}{\pi}\int_{\mathbb C}Q(|z_-|^2) \ln\left(Q(|z_-|^2)\right) d^2 z_- \right\}\cr
&=& \left[1 + \ln\left(\frac{1}{e^{\beta \hbar \omega_c } -1}\right)\right]\times
\left[1 + \ln\left(\frac{1}{e^{\beta \hbar \omega^{*}_c } -1}\right)\right] 
\end{eqnarray}
}

where the following variables changes $u_+ = \frac{1}{\bar{n}+1}r^2_+$ and $u_-= \frac{1}{{\bar{n}}^{*}+1}r^2_-, $ have been performed.

Thereby,
{\small{
\begin{eqnarray}
W 
= \left[1-\ln \left(1-e^{- \beta \hbar \omega_c }\right)\right]\left[1-\ln \left(1-e^{- \beta \hbar \omega^{*}_c }\right)\right].
\end{eqnarray}
}}
$\hfill{\square}$

\subsection{A bit on the Landau diamagnetism}

By  following the approach of \cite{omer-jellal}, 
 we   take the expression of the partition function for a cylindrical body of length $L$, 
radius $R$ and of volume $V, $ which is oriented along the $z$-direction  in noncommutative
 coordinates,  to be

{\small{
\begin{equation}
Z  = \frac{V}{\lambda^{3}} \frac{\beta \hbar  \omega^{*}_{c}}{2} \frac{1}{\sinh\left\{\beta \hbar  \omega^{*}_{c}/2 \right\}}
\end{equation}}}
where $\lambda = (2\pi \hbar^{2}/M)^{1/2}$ is the thermal wavelength.

Taking the expressions  for the free energy, 
the magnetization and the susceptibility in the standard definitions, i.e.    
{\small{\begin{equation}
F = -\frac{n}{\beta} \ln \mathcal Z,   \qquad  \mathcal M = -\frac{\partial F}{\partial B},  \qquad 
\chi = \frac{1}{n}\frac{\partial \mathcal M}{\partial B}, 
\end{equation}}}
respectively, we get  the modified quantities for the exotic Landau model as follows: 

{\small{\begin{eqnarray}\label{freenergy}
F  &=& -\frac{n}{\beta}\left[\ln \frac{V}{\lambda^{3}} + \ln \frac{\beta \hbar \omega_c}{2}(1-eB\theta) - \ln \left(
\sinh\left\{\frac{\beta \hbar \omega_c}{2}(1-eB\theta)\right\}\right)\right],\cr
\mathcal M  &=& n \frac{\hbar e}{Mc}(1-2eB\theta)\left[\frac{1}{\beta \hbar \omega_c(1-eB\theta)} - 
\frac{1}{2}\coth\left\{\frac{\beta \hbar \omega_c}{2}(1-eB\theta)\right\} \right],
\end{eqnarray}}}
 
{\small{\begin{eqnarray}\label{suscep}
\chi &=& - 2\frac{\hbar e}{Mc}e\theta \left[\frac{1}{\beta \hbar \omega_c(1-eB\theta)} 
- \frac{1}{2}\coth\left\{\frac{\beta \hbar \omega_c}{2}(1-eB\theta)\right\}\right] -\left(\frac{\hbar e}{Mc}\right)^2 
\beta (1-2eB\theta)^2 \times\cr
&&\times \left[\frac{1}{\left(\beta \hbar \omega_c(1-eB\theta)\right)^2} 
+ \frac{1}{4}\left(1-\coth^2\left\{\frac{\beta \hbar \omega_c}{2}(1-eB\theta)\right\}\right)\right].
\end{eqnarray}}}

In the high temperature limit, $\beta \ll 1$ i.e. $x=  \beta \hbar \omega_c(1-eB\theta) \ll 1$, we obtain

{\small{\begin{equation}\label{diamagn}
\chi = -\frac{1}{3}\left(\frac{\hbar e}{2Mc}\right)^2 \beta [1+6\kappa + 6\kappa^2]
\end{equation}}}

 with 
$\kappa = -eB\theta$. Since $1-eB\theta >0$ in our model, taking the noncommutativity parameter $\theta$ positive such that 
 $eB\theta < 1$,   the system is then  diamagnetic, except for the values 
$-0.8 < \kappa < -0.2$ where $\chi$ becomes positive.

 The free energy, magnetization and susceptibility derived in (\ref{freenergy}) and (\ref{diamagn}), 
respectively, are found to be standard ones when
 
\begin{enumerate}

\item $\kappa= -eB\theta \ll 1, $ i.e 
{\small{\begin{equation} 
F \simeq -\frac{n}{\beta}\left[\ln \frac{V}{\lambda^{3}} + \ln \frac{\beta \hbar \omega_c}{2} - \ln \left(
\sinh\left\{\frac{\beta \hbar \omega_c}{2}\right\}\right)\right], \;
\mathcal M \simeq  n \frac{\hbar e}{Mc}\left[\frac{1}{\beta \hbar \omega_c} - 
\frac{1}{2}\coth\left\{\frac{\beta \hbar \omega_c}{2}\right\} \right];
\end{equation}}}

\item $\kappa= -eB\theta  \rightarrow 0, $ i.e
{\small{\begin{equation}
\chi  \simeq  -\frac{1}{3}\left(\frac{\hbar e}{2Mc}\right)^2 \beta  
\end{equation}}}
which is the usual Landau diamagnetism.

\end{enumerate}

 \section{Density operator approach in CS related to  the harmonic oscillator thermal state}
 
We start by sketching key ingredients from \cite{bratelli, alietal, takesaki} as needed for this section.

\bdfn

Consider the unitary operator 
$U(x,y)$ on $\mathfrak H$ given by 
{\small{
\begin{eqnarray}
(U(x,y)\Phi)(\xi) =  e^{-i x \left(\xi - y/2 \right)}\Phi\left(\xi -y \right),
\end{eqnarray}}}
$x,y, \xi \in \mathbb R$, where $U(x,y) = e^{-i (xQ + y P)}$, with $[Q, P] = i \mathbb I_{\tilde{\mathfrak H}}$, and the Wigner transform  given by
{\small{\begin{equation}{\label{map1}}
\mathcal W: \mathcal B_{2}(\mathfrak H) \rightarrow L^{2}(\mathbb R^{2}, dxdy) =   \tilde{\mathfrak H},\, \,(\mathcal WX)(x,y) = \frac{1}{(2\pi)^{1/2}}Tr[(U(x,y))^{*}X],  
\end{equation}
}}
where $X \in \mathcal B_{2}(\mathfrak H), x,
 y \in \mathbb R$. $\mathcal W$ is unitary. 
\edfn

\bdfn

Let $\mathcal B_{2}(\mathfrak H)$, $\mathcal B_{2}(\mathfrak H) \subset \mathcal L(\mathfrak H)$ 
the set of all bounded operators on $\mathfrak H$,
be the Hilbert space of Hilbert-Schmidt 
operators on $\mathfrak H = L^{2}(\mathbb R)$, with the scalar product 
{\small{
\begin{equation}\label{scalprod00}
\langle X|Y \rangle_{2} = Tr[X^{*}Y] = \sum_{k=0}^{\infty}\langle \Phi_{k}|X^{*}Y\Phi_{k}\rangle,
\end{equation} 
}}
where $\{\Phi_{k}\}^{\infty}_{k=0}$ is an orthonormal basis of $\mathfrak H$.

\edfn

$\mathcal B_{2}(\mathfrak H)\simeq \mathfrak H \otimes \bar{\mathfrak H}$ (where $\bar{\mathfrak H}$ denotes the dual of $\mathfrak H$), 
and basis vectors of $\mathcal B_{2}(\mathfrak H)$ are  given by
{\small{
\begin{equation}{\label{basis01}}
\Phi_{nl} := |\Phi_{n}\rangle \langle \Phi_{l}|, \quad n,  l = 0,1,2,\dots, \infty.
\end{equation}
 }}
\bdfn

Let  $A$ and  $B$ two operators on  $\mathfrak H$. The operator 
$A \vee B$ is such that 
{\small{
\begin{equation}\label{modunit02}
A \vee B (X) = AXB^{*}, \; X \in B_2(\mathfrak H).
\end{equation}
}}
\edfn

For $A$ and $B$,  both bounded operators,  $A \vee B$ defines a linear operator on $B_2(\mathfrak H)$.

{\bf Kubo-Martin-Schwinger (KMS) state}

 Let $\alpha_i, \, i=1, 2, \dots, N$ be a sequence of non-zero, positive numbers, satisfying :  $\sum_{i=1}^{N}\alpha_i = 1. $
 Let 
 {\small{
\begin{equation}\label{vectphi00}
\Phi = \sum^{N}_{i=1} \alpha^{\frac{1}{2}}_i \mathbb P_i = \sum^{N}_{i=1} \alpha^{\frac{1}{2}}_i X_{ii} \in \mathcal{B}_2(\mathfrak{H}) \quad \mbox{with} \quad X_{ii} = |\zeta_i\rangle \langle \zeta_i|.
\end{equation}
}}
Then, we have the following properties:

\begin{enumerate}
\item

\bpro

 $\Phi$ defines a vector state $\varphi$ on the von Neumann algebra $\mathfrak{A}_l$ 
 corresponding to the operators given with $A$ in the left of the identity operator 
 $I_{\mathfrak H}$ on $\mathfrak H, $ i.e.,  $\mathfrak{A}_l= \left\{A_{l} = A \vee I| A \in \mathcal L(\mathfrak H)\right\}.$
\epro

{\bf Proof.}  See \cite{aremuaetal-HSchm}.

$\hfill{\square}$

\item

\bpro

 The state $\varphi$ is faithful and normal.
\epro

{\bf Proof.} See \cite{aremuaetal-HSchm}.

$\hfill{\square}$

\item

\bpro

 The vector $\Phi$ is cyclic and separating for $\mathfrak{A}_l.$
\epro

{\bf Proof.} See \cite{aremuaetal-HSchm}.

$\hfill{\square}$

 The fact that $\Phi$ is separating for $\mathfrak{A}_l$ is obtained through the relation 

 {\small{
\begin{equation}
(A \vee I)\Phi = (B \vee I)\Phi \Longleftrightarrow A \vee I = B \vee I, \quad \forall A, B \in \mathfrak{A}_l.
\end{equation}
}}

{\bf Proof.}  See \cite{aremuaetal-HSchm}.

$\hfill{\square}$

\end{enumerate}

 In the same way, $\Phi$ is also cyclic for $\mathfrak{A}_r = \left\{A_{r} = I \vee A| A \in \mathcal L(\mathfrak H)\right\},$ which corresponds to the operators given with $A$ in the right of the identity operator $I_{\mathfrak H}$ on $\mathfrak H, $ hence separating for $\mathfrak{A}_r, $ i.e. 
$(I \vee A)\Phi = (I \vee B)\Phi \Longleftrightarrow I \vee A = I \vee B.$

\subsection{Thermal state}
Here, we give two examples  of  thermal states as known from  the literature. 
For more details, see  \cite{tomita, takesaki, bratelli, alietal, aremuaetal-HSchm}.

\begin{enumerate}
\item
Let $\alpha_i, i= 1,2,\dots, N$ be a sequence of non-zero, positive numbers, satisfying
$\sum_{i=1}^{N}\alpha_i = 1.$  Then the thermal state is defined as:
{\small{
\begin{equation}
\Phi := \sum_{i=1}^{N}\alpha^{\frac{1}{2}}_i \mathbb P_i = 
\sum_{i=1}^{N}\alpha^{\frac{1}{2}}_i 
X_{ii}  \in \mathcal B_2(\mathfrak H),
\end{equation}
}}
where $\mathbb P_i = X_{ii}= |\zeta_i\rangle \langle \zeta_i|$.
 
\item The thermal equilibrium state $\Phi$ at inverse temperature $\beta, $
corresponding to the harmonic oscillator  Hamiltonian $H_{OSC} = \frac{1}{2}(P^2 + Q^2), $ with 
$H_{OSC} \phi_n = \omega(n + \frac{1}{2})\phi_n, n= 0, 1, 2, \dots, $ where the density matrix is
{\small{
\begin{equation}
\rho_{\beta} = \frac{ e^{-\beta H_{\mbox{osc}}}}{Tr\left[ e^{-\beta
H_{\mbox{osc}}}\right]} = (1 -  e^{-\omega \beta})\sum_{n=0}^{\infty}
 e^{-n \omega \beta}|\phi_n\rangle \langle \phi_n|, \quad  \mbox{Tr}[e^{-\beta H_{OSC}}] = \frac{e^{-\frac{\beta \omega}{2}}}{1- e^{-\beta \omega}},
\end{equation}
}}
is 
{\small{
\begin{equation}
\Phi = \left[1-e^{-\omega\beta}\right]^{\frac{1}{2}}\sum_{n=0}^{\infty}e^{-\frac{n}{2} \omega\beta}|\phi_n\rangle \langle \phi_n|.
\end{equation}
}}

\end{enumerate}
 
\subsection{Coherent states built from the harmonic oscillator thermal state}

Take the cyclic vector $\Phi$ of the von Neumann algebra $\mathfrak A_1$ 
generated by the unitary 
operator
{\small{
\begin{equation}\label{unitop}
U_1(x,y) = \mathcal{W}\left[U(x,y) \vee I_{\mathfrak H}\right]
\mathcal{W}^{-1},\quad \mbox{with}\; \; \mathcal{W} \;\;  \mbox{given  in (\ref{map1})}  
\end{equation}
}}
such that $\Phi=\Phi_\beta$ with the $\lambda_n$ corresponding  to the thermal state $\Phi_\beta$
{\small{
\begin{equation}\label{thermalst00}
\Phi_{\beta} = \left[1 -  e^{-\omega \beta}\right]^{\frac{1}{2}}
\sum_{n=0}^{\infty} e^{-n\frac{\omega \beta}{2}}\Psi_{nn}, \; \mbox{i.e.},\;
\lambda_n = (1- e^{-\omega \beta}) e^{-n \omega \beta}.
\end{equation}
}}

The CS, denoted  $|z,\bar{z},\beta\rangle^{\mbox{\tiny{KMS}}}, $ built from  the
thermal state $\Phi_{\beta}$, are given by
{\small{
\begin{equation}
|z,\bar{z},\beta\rangle^{\mbox{\tiny{KMS}}} = U_1(z)|\Phi_{\beta}\rangle =
 e^{zA^{\dag}_1-\bar{z}A_1}|\Phi_{\beta}\rangle.
\end{equation}
}}
with $U_1(z):= U_1(x,y) = e^{zA^{\dag}_1-\bar{z}A_1},$ where the actions of the annihilation and creation operators, 
$A_1$ and $A^{\dag}_1$ are given by 
{\small{
\begin{equation}
 A^{\dag}_1|\Psi_{nl}\rangle=\sqrt{n+1} \Psi_{n+1l}, \qquad  A_1|\Psi_{nl}\rangle=\sqrt{n} \Psi_{n-1l}.
\end{equation}
}}

\bpro\label{unitopvect00}
From the fact that  the states $\phi_i,\; i=0,1,2,\cdots,\infty,$
form a basis of  $\mathfrak H = L^{2}(\mathbb R)$, the following equalities 
{\small{
\begin{equation}
U_1(x,y)\Phi_{\beta} = (2\pi)^{\frac{1}{2}}\sum_{i,j=0}^{\infty}
\lambda^{\frac{1}{2}}_i\overline{\Psi_{ji}(x,y)}\Psi_{ji}, \quad 
U_1(x,y)^{*}\Phi_{\beta} =(2\pi)^{\frac{1}{2}} \sum_{i,j=0}^{\infty}\lambda^{\frac{1}{2}}_j
\Psi_{ji}(x,y)\Psi_{ij}
\end{equation}
}}
hold.
\epro

{\bf Proof.}  See \cite{aremuaetal-HSchm}.

$\hfill{\square}$

\bpro (\cite{alietal})

From the isometry $\mathcal W, $ the CS $|z,\bar{z},\beta\rangle^{\mbox{\tiny{KMS}}}$ satisfy the resolution  of the identity condition
{\small{
\begin{eqnarray}\label{KMSresolv}
\frac{1}{2\pi}\int_{\mathbb C}|z,\bar{z},\beta\rangle^{\mbox{\tiny{KMS}}}
{^{\mbox{\tiny{KMS}}}}\langle z,\bar{z},\beta|dxdy = I_{\tilde{\mathfrak{H}}}, \;\;\; \tilde{\mathfrak{H}} = L^2(\mathbb R^2, dxdy).
\end{eqnarray}
}}
\epro

{\bf Proof.}  By definition of the states $|z,\bar{z},\beta\rangle^{\mbox{\tiny{KMS}}}$ and Proposition \ref{unitopvect00},
 we have:
 {\small
\begin{eqnarray}\label{KMSvect00}
|z,\bar{z},\beta\rangle^{\mbox{\tiny{KMS}}}  
=(2\pi)^{\frac{1}{2}}\sum_{i,j=0}^{\infty}\lambda_i^{\frac{1}{2}}\overline{\Psi_{ji}(x,y)}|\Psi_{ji}\rangle, \, \, \, 
^{\mbox{\tiny{KMS}}}\langle z,\bar{z},\beta|   = (2\pi)^{\frac{1}{2}}
\sum_{l,k=0}^{\infty}\langle\Psi_{lk}|\lambda_k^{\frac{1}{2}}\Psi_{lk}(x,y). 
\end{eqnarray}
}
Thereby
{\small{
\begin{eqnarray}\label{onyx}
|z,\bar{z},\beta\rangle^{\mbox{\tiny{KMS}}}{^{\mbox{\tiny{KMS}}}}\langle z,\bar{z},\beta|
= 2\pi\sum_{i,j=0}^{\infty}\sum_{l,k=0}^{\infty}\lambda^{\frac{1}{2}}_i
\lambda^{\frac{1}{2}}_k
\overline{\mathcal{W}\phi_{ji}(x,y)}\mathcal{W}\phi_{lk}(x,y)
|\Psi_{ji}\rangle\langle\Psi_{lk}|.
\end{eqnarray}
}}
 By integrating the two members of the Eq.(\ref{onyx}) over 
$\mathbb C$, we get by using the Wigner map $\mathcal W, $
{\small
\begin{eqnarray}
\frac{1}{2\pi}\int_{\mathbb C}|z,\bar{z},\beta\rangle^{\mbox{\tiny{KMS}}}{^{\mbox{\tiny{KMS}}}}\langle z,\bar{z},\beta|dxdy 
&=& \sum_{i,j=0}^{\infty}\sum_{l,k=0}^{\infty}\lambda^{\frac{1}{2}}_i\lambda^{\frac{1}{2}}_k |\Psi_{ji}\rangle \langle\Psi_{lk}| 
\int_{\mathbb R^2}\overline{\mathcal{W}\phi_{ji}(x,y)}\mathcal{W}\phi_{lk}(x,y)dxdy \cr
&=& \sum_{i,j=0}^{\infty}\sum_{l,k=0}^{\infty}\lambda^{\frac{1}{2}}_i\lambda^{\frac{1}{2}}_k|\Psi_{ji}\rangle \langle\Psi_{ji}| \delta_{lj}\delta_{ki}\cr
&=& I_{\tilde{\mathfrak H}}.
\end{eqnarray}
}
$\hfill{\square}$

\bpro\label{KMScomponents}
The components of the KMS CS $|z,\bar{z},\beta\rangle^{\mbox{\tiny{KMS}}}$ in the states $|\phi_n\rangle$ are derived as 
{\small
\begin{eqnarray}
|\langle \phi_n|z,\bar{z},\beta\rangle^{\mbox{\tiny{KMS}}}|^{2} &=& 
\left[1 -  e^{-\omega \beta}\right]\sum_{s,t=0}^{\infty}\sum_{i,j=0}^{\infty}e^{-(t+j)\frac{\omega \beta}{2}}\times \cr
&& \left\{2\pi\, \langle \phi_n|\Psi_{ij}\rangle \overline{\mathcal{W}(|\phi_{i}\rangle \langle \phi_{j}|)(x,y)}
\mathcal{W}(|\phi_{s}\rangle \langle \phi_t|) (x,y) \langle\Psi_{st}| \phi_n\rangle \right\}.
\end{eqnarray}
}

\epro

{\bf Proof.}  According to (\ref{thermalst00}), it follows that
{\small
\begin{eqnarray}
|\langle \phi_n|z,\bar{z},\beta\rangle^{\mbox{\tiny{KMS}}}|^{2} 
 &=& \left[1 -  e^{-\omega \beta}\right]\sum_{s,t=0}^{\infty}\sum_{i,j=0}^{\infty}e^{-(t+j)\frac{\omega \beta}{2}}\times \cr
&& \left\{2\pi\, \langle \phi_n|\Psi_{ij}\rangle \overline{\mathcal{W}(|\phi_{i}\rangle \langle \phi_{j}|)(x,y)}
\mathcal{W}(|\phi_{s}\rangle \langle \phi_t|) (x,y) \langle\Psi_{st}| \phi_n\rangle \right\}. 
\end{eqnarray}
}
$\hfill{\square}$

Next, from the Proposition \ref{KMScomponents}, the $Q$-Husimi distibution is performed  as
{\small
\begin{eqnarray}\label{KMSqhusimi}
{^{\mbox{\tiny{KMS}}}}\langle z,\bar{z},\beta|\rho|z,\bar{z},\beta\rangle^{\mbox{\tiny{KMS}}}
&=& 
\left[1 -  e^{-\omega \beta}\right]^2\sum_{n=0}^{\infty}
\sum_{s,t=0}^{\infty}\sum_{i,j=0}^{\infty}e^{-n\omega \beta} e^{-(t+j)\frac{\omega \beta}{2}}\times \cr
&& \left\{2\pi\, \langle \phi_n|\Psi_{ij}\rangle \overline{\mathcal{W}(|\phi_{i}\rangle \langle \phi_{j}|)(x,y)}
\mathcal{W}(|\phi_{s}\rangle \langle \phi_t|) (x,y) \langle\Psi_{st}| \phi_n\rangle \right\}. \nonumber \\
\end{eqnarray}
}

The $P$-distribution function is found from the following result:

\bpro
From the Glauber-Sudarshan $P$-distribution function   in the KMS CS $|z,\bar{z},\beta\rangle^{\mbox{\tiny{KMS}}}$, 
{\small{
\begin{equation}
\rho = \frac{1}{2\pi}
\int_{\mathbb C} dx dy P(|z|^2)|z,\bar{z},\beta\rangle^{\mbox{\tiny{KMS}}}{^{\mbox{\tiny{KMS}}}}\langle z,\bar{z},\beta|
\end{equation}
}}
the diagonal elements of the  normalized density operator $\rho$ in the basis of the states $\{|\phi_n\rangle\}_{n=0}^{\infty}$, 
by using the thermal 
state  (\ref{thermalst00}) and the resolution of the identity (\ref{KMSresolv}), are linked to the $P$-distribution function through the 
following equality

{\small{
\begin{equation}
\langle \phi_n|\rho|\phi_n\rangle = \frac{1}{\bar{n}_0+1}\left(\frac{\bar{n}_0}{\bar{n}_0+1}\right)^{n} = 
\sum_{k,l=0}^{\infty}
  \lambda_k 
|\left\langle \phi_n \right|\Psi_{lk} \rangle|^{2}
\left\{\int_{\mathbb C}  \frac{d^2 z}{\pi} P(|z|^2)
\right\} 
\end{equation}
}}
where $\bar{n}_0 = \left[1- e^{-\omega \beta}\right]^{-1} $ is the thermal occupancy for the harmonic oscillator  
 with the angular frequency $\omega$.
\epro

{\bf Proof.}  Let us start with the definition of Glauber-Sudarshan $P$-distribution function \cite{glauber, sudarshan}, 
{\small{
\begin{equation}
\rho = \frac{1}{2\pi}
\int_{\mathbb C} dx dy P(|z|^2)|z,\bar{z},\beta\rangle^{\mbox{\tiny{KMS}}}{^{\mbox{\tiny{KMS}}}}\langle z,\bar{z},\beta|
\end{equation}
}}
such that we get
{\small
\begin{eqnarray}
\langle \phi_n|\rho|\phi_n \rangle 
&=& \frac{1}{2\pi}\int_{\mathbb C} dx dy P(|z|^2)|\langle \phi_n|z,\bar{z},\beta\rangle^{\mbox{\tiny{KMS}}}|^{2}   \cr
&=& \left[1 -  e^{-\omega \beta}\right]\sum_{s,t=0}^{\infty}\sum_{i,j=0}^{\infty}e^{-(t+j)\frac{\omega \beta}{2}}\times \cr
&& \left\{\int_{\mathbb C}  P(|z|^2)|\langle \phi_n|\Psi_{ij}\rangle \overline{\mathcal{W}(|\phi_{i}\rangle \langle \phi_{j}|)(x,y)}
\mathcal{W}(|\phi_{s}\rangle \langle \phi_t|) (x,y) \langle\Psi_{st}| \phi_n\rangle  dx dy\right\}.
\end{eqnarray}
}
Then,  introducing the differential element of area in the $z$ plane \cite{glauber}, 
 $\frac{d^2 z}{\pi} = d(Re z)d(Im z) = \frac{dx dp}{2\pi} (\hbar = 1), $  it follows that
{\small
\begin{eqnarray}
  \int_{\mathbb C}  \frac{d^2 z}{\pi} \langle \phi_n|\rho|\phi_n\rangle  
&=& \sum_{k,l=0}^{\infty}\sum_{i,j=0}^{\infty}\lambda_i^{\frac{1}{2}}\lambda_k^{\frac{1}{2}}  \left\{\int_{\mathbb C}  \frac{d^2 z}{\pi} P(|z|^2)
\right\} \left\{\delta_{lj} \delta_{ik} \right\}\langle \Psi_{lk}|\phi_n\rangle\langle \phi_n|\Psi_{ji}\rangle. 
\end{eqnarray}
}

Thus 
 
{\small{
\begin{eqnarray}\label{Pdistrib00}
\langle \phi_n|\rho|\phi_n\rangle = \frac{1}{\bar{n}_0+1}\left(\frac{\bar{n}_0}{\bar{n}_0+1}\right)^{n} = 
\sum_{k,l=0}^{\infty}
  \lambda_k 
|\left\langle \phi_n \right|\Psi_{lk} \rangle|^{2}
\left\{\int_{\mathbb C}  \frac{d^2 z}{\pi} P(|z|^2)
\right\}
\end{eqnarray}
}}
where the thermal occupancy for the harmonic oscillator
$\bar{n}_0 = \left[1- e^{-\omega \beta}\right]^{-1}$ with the angular frequency $\omega$,   has been introduced.

$\hfill{\square}$

Using the $Q$-Husimi distribution expression (\ref{KMSqhusimi}), the Wehrl entropy is deduced as 
{\small{
\begin{equation}
W = \int_{\mathbb C}  \frac{d^2 z}{\pi} \, {^{\mbox{\tiny{KMS}}}}\langle z,\bar{z},\beta|\rho|z,\bar{z},\beta\rangle^{\mbox{\tiny{KMS}}}\, 
\ln{\left\{{^{\mbox{\tiny{KMS}}}}\langle z,\bar{z},\beta|\rho|z,\bar{z},\beta\rangle^{\mbox{\tiny{KMS}}}\right\}}.
\end{equation}
}}

\section{Concluding remarks}

In this work, the definition of the density operator on quantum states in
Hilbert spaces and some of its features relevant in thermodynamics and information-theoretical entropy calculations have been  provided. 
As application,  the  physical 
model describing an electron in a magnetic
field has been studied. The  exotic Landau problem in noncommutative plane has been investigated and 
 the related CS have been constructed. In addition,  
 the quantum model for which 
 modular structures emerging   for two underlying von Neumann algebras have been provided, has been revisited. 
 The resolution of the identity satisfied by the CS
 built out of Kubo-Martin-Schwinger (KMS)
state has been achieved.  Thanks to the completeness relations verified by the CS in these two examples, 
the  thermodynamics has been discussed, using  the diagonal $P$-representation of the
density operator  in the constructed
CS and in the Hilbert space basis. Besides, the $Q$-Husimi
distribution and the Wehrl entropy
have been  determined. 
 
\section{Acknowledgements}
The authors thank the CIRM workshop {\it Coherent states and their applications:
 A contemporary panorama} organizers.
This work is partially supported by the ICTP through the  OEA-ICMPA-Prj-15. The ICMPA-UNESCO Chair  is in partnership with the Daniel
Iagolnitzer Foundation (DIF), France.

\end{document}